\documentstyle[12pt,a41,epsfig]{article}

\newcommand{\bea}{\begin{eqnarray}}
\newcommand{\bq}{\begin{equation}}
\newcommand{\eea}{\end{eqnarray}}
\newcommand{\eq}{\end{equation}}

\newcommand\epsi{\varepsilon}

\newcommand\xx{\tilde{x}}

\newcommand\T{{\sf T}}
\begin{document}
\noindent
\sloppy
\thispagestyle{empty}
\begin{flushleft}
DESY 00--005 \hfill
{\tt hep-ph/0002071}\\
UNIGRAZ UTP 11-01-00\\
January 2000
\end{flushleft}
%
\vspace*{\fill}
\begin{center}
{\LARGE\bf  On the Structure of the Virtual Compton}\\
\vspace{2mm}
{\LARGE\bf
Amplitude in the
Generalized Bjorken Region:}\\
\vspace{2mm}
{\LARGE\bf Integral Relations}\\

\vspace{2cm}
\large
Johannes Bl\"umlein$^a$  and
Dieter Robaschik$^{a,b}$
\\
\vspace{2em}
\normalsize
{\it $^a$~Deutsche Elektronen--Synchrotron,\\
DESY -- Zeuthen, Platanenallee 6, D--15735 Zeuthen, Germany}
\\
\vspace{2em}
{\it $^b$~Institut f\"ur Theoretische Physik,
 Karl--Franzens--Universit\"at Graz,}\\
{\it  Universit\"atsplatz  5, A--8010 Graz, Austria} \\
\end{center}
\vspace*{\fill}
%
\begin{abstract}
\noindent
A systematic study is performed of the structure of the twist--2
contributions to the virtual Compton amplitude in deep--inelastic 
non--forward scattering for general spin states in the non--local
light cone expansion. The gauge invariance is proven in this approach
for the twist--2 contributions for non--forward scattering. Relations 
are derived between the contributing non--forward amplitudes.
\end{abstract}
\vspace*{\fill}
\newpage
\section{Introduction}
\renewcommand{\theequation}{\thesection.\arabic{equation}}
\setcounter{equation}{0}
\label{sec-1}

\vspace{1mm}
\noindent
The Compton amplitude for the scattering of a virtual photon off a hadron
\begin{equation}
\label{eqpro}
\gamma^*_1 + p_1 \rightarrow \gamma_2^{(*)'} + p_2
\end{equation}
provides one of the basic tools to understand the short--distance
behavior of the nucleon and to test Quantum Chromodynamics (QCD) at
large space--like virtualities. In this kinematic regime the Compton
amplitude is dominated by the singularities of the light--cone and can be
described by the light--cone expansion~\cite{LC1,LC2} in terms of the
contributions of different twist~\cite{TW}. Many investigations were
devoted to the process of deep--inelastic forward scattering in the past, 
for which the leading, next--to--leading order, and partly the 3--loop
order corrections were calculated~\cite{RC}. There the absorptive part of
the Compton amplitude, the hadronic tensor $W_{\mu\nu}$, is described
by the four independent structure functions $F_1(x,Q^2), F_2(x,Q^2),
g_1(x,Q^2)$ and $g_2(x,Q^2)$~\cite{DER}.\footnote{In the case of general 
electroweak currents eight structure functions contribute in the massless
limit, cf.~\cite{BK,BT}.}

A more general view is obtained studying the non--forward 
process~(\ref{eqpro}) without restricting it to the absorptive 
contributions only. In this more general kinematics a systematic 
description of a whole class of physical processes in the deep--inelastic
region can be given and common properties of these processes can be
derived. Moreover, relations 
obtained for forward scattering become more transparent under this 
generalized point of view. It is the aim of this paper to calculate the 
Compton amplitude for non--forward scattering in this kinematic range
in lowest order in QCD. General spin states are considered which allows
us to study besides the unpolarized contributions also those due to
target polarization. The present analysis is devoted to the study of the
twist--2 terms. The twist--decomposition of the Compton amplitude in
the generalized Bjorken region is necessary since the scaling violations 
of the respective amplitude functions differ due to the set of
the contributing operators and their anomalous dimensions~\cite{ADI}.
The different contributions to the Compton amplitude are described by 
non--perturbative amplitude functions, which can be measured 
experimentally in the various non--forward processes.
Analogously to the case of forward scattering some of the amplitude
functions are experimentally easier accessible than others\footnote{
Both the measurement of the unpolarized longitudinal structure function 
$F_L(x,Q^2) = F_2(x,Q^2) - 2x F_1(x,Q^2)$~\cite{FL} and the polarized
structure function $g_2(x,Q^2)$~\cite{G2} are experimentally much more
demanding than the measurement of the structure functions $F_2(x,Q^2)$ 
and $g_1(x,Q^2)$.}. Therefore it is important to know the relations
between the different amplitude functions due to which predictions can be
made for experiment. One of the main objectives of this investigation 
is to derive these relations for the twist--2 parts of the amplitude 
functions.

The paper is organized as follows. In Section~2 the twist--2
contributions to the Compton amplitude are calculated. We apply the
non--local light--cone expansion~\cite{NLC,BGR} to express the operator
products. The non--forward expectation values are calculated for general
spin states and the corresponding Lorentz--structure is derived. A 
helicity basis of both (virtual) photons is constructed (Section~3) which
is subsequently used to determine the twist--2 contributions to the 
operator matrix elements. The light--cone expansion does not a priori
lead to an explicit gauge--invariant representation of the different
contributions to the Compton amplitude. Since the Compton amplitude 
itself is gauge invariant, the expansion has to be formulated in such a 
way that the respective contributions under discussion form 
gauge--invariant sub--sets. In Section~4 we show that this is the case
for the twist--2 contributions for non--forward scattering.
The helicity projections of the Compton amplitude are discussed
in detail in Section~5. In Section~6 we derive the relations between the
twist--2 contributions to the different amplitude functions, which are
generalizations of the relations known in the case of forward 
scattering~\cite{CG,WW}. 
Section~7 contains the conclusions.
\section{The Compton Amplitude}
\renewcommand{\theequation}{\thesection.\arabic{equation}}
\setcounter{equation}{0}
\label{sec-2}
The Compton amplitude for the general case of
non--forward scattering is given by
\begin{equation}
\label{COMP}
T_{\mu\nu}(p_+,p_-,q)
= i \int d^4x \,e^{iqx}\,
\langle p_2, S_2\,|T (J_{\mu}(x/2) J_{\nu}(-x/2))|\,p_1,
S_1\rangle~.
\end{equation}
Here,
\begin{eqnarray}
p_+ &=& p_2 + p_1,~~~~~~~~~~p_- = p_2 - p_1 = q_1 - q_2, \\
q   &=& \hbox{\large $\frac{1}{2}$}
        \left(q_1 + q_2\right),~~~~~~~~p_1 + q_1 = p_2 + q_2~,
\end{eqnarray}
where $q_1~(q_2)$ and $p_1~(p_2)$ denote the four--momenta of the
incoming~(outgoing) photon and hadron, respectively, and $S_{1}, S_{2}$
are the spins of the initial-- and final--state hadron.
Representations of the Lorentz--structure of the Compton amplitude were
given in Refs.~\cite{LR}. In the following we will consider the
Compton amplitude in the {\it generalized Bjorken region} only which
is defined by the conditions
\begin{eqnarray}
\label{gBr1}
\nu =  qp_+ \longrightarrow \infty, \qquad
- q^2 \longrightarrow \infty~,
\end{eqnarray}
keeping the variables
\begin{eqnarray}
\label{gBr2}
\xi  = - \frac{q^2 }{qp_+}, \qquad
\eta = \frac{qp_-}{qp_+} = \frac{q_1^2 - q_2^2}{2\nu}
\end{eqnarray}
fixed. The subsequent analysis will be performed demanding furthermore
that the vector $q_1$ is spacelike and the vector $q_2$ can be
space--, light--, or time--like,
where
\begin{eqnarray}
q_1 &=& q + \hbox{\large $\frac{1}{2}$}p_- \\
q_2 &=& q - \hbox{\large $\frac{1}{2}$}p_-~.
\end{eqnarray}
In the generalized Bjorken region the Compton amplitude is dominated by
the light--cone singularities. It is therefore possible to apply the
light--cone expansion for its representation. In the following we use
the non--local light-cone expansion~\cite{NLC,BGR}, which is a 
summed--up form of the local light--cone expansion with respect to the
spin indices. The respective expressions in the local light--cone
expansion can be obtained from the former one by a Taylor expansion.
In this way more compact representations can be obtained, 
cf.~section~\ref{sec-6}.

In this paper we study the twist--2 contributions to the 
non--forward Compton amplitude. They are obtained from the expectation
values of the non-local twist--2 light cone operators, 
cf.~\cite{BGR,GLR}. Here the notion of twist is used in its original
form as canonical dimension - spin~\cite{TW} for the local operators
which are summed up to the non--local operators. In calculating the
non--forward operator expectation values it turns out that the 
twist--decomposition performed for the operators is not necessarily
complete in the case of the expectation values unlike the case for
forward scattering. The emergence of new hadronic mass scales, such as
$p_+.p_-$ or also of off--shell terms $\propto
p_+^2, p_-^2$ open  the possibility that expectation values of operators
of lower twist mix with expectation values of operators of higher twist
by virtue of these terms, cf. e.g.~\cite{BN}. The light cone expansion in
the
non--forward case bears therefore additional complications to be dealt
with. The impact of the hadronic mass scales $p_+^2, 
p_-^2$ and $p_+.p_-$,
which is of relevance for the contributions beyond leading twist, can in
general not be dealt with as target mass-- or final--state mass
corrections only~cf.~\cite{MIR,BT} for a discussion. Furthermore,
the scales $p_+^2$ and $p_-^2$  dependend on the factorization
of non--perturbative quantities as the distribution
amplitudes and are arbitrary in this sense. In size they are comparable
to the mass scales of the expectaion values of higher twist operators.
\subsection{Operator Structure}
\renewcommand{\theequation}{\thesection.\arabic{equation}}
\label{sec-2.1}

\vspace{1mm}
\noindent
The product of the two currents 
\begin{eqnarray}
\hat{T}_{\mu\nu}(x)  =
i RT \left[J_\mu\left(\frac{x}{2}\right)
J_\nu\left(-\frac{x}{2}\right) S \right]
\end{eqnarray}
is then given by~\footnote{Here and in the following, 
the normal product symbols for the operator products will be omitted.}
\begin{eqnarray}
\label{Tmunu}
 \hat{T}^{\mu\nu}(x) = 
 -e^2 \frac{\tilde x^\lambda}{2 \pi^2 (x^2-i\epsilon)^2}
 RT
 \left[
\overline{\psi}
\left(\frac{\tilde x}{2}\right)
\gamma^\mu \gamma^\lambda \gamma^\nu \psi
\left(-\frac{\tilde x}{2}\right)
- \overline{\psi}
\left(-\frac{\tilde x}{2}\right)
\gamma^\mu \gamma^\lambda \gamma^\nu \psi
\left(\frac{\tilde x}{2}\right)
\right] S~.
\end{eqnarray}
$\tilde x$ denotes a light--like vector corresponding to $x$,
\begin{eqnarray}
\label{xtil}
\tilde x = x + \frac{\zeta}{\zeta^2}\left[ \sqrt{x.\zeta^2 - x^2 \zeta^2}
- x.\zeta\right]~,
\end{eqnarray}
and $\zeta$ is a subsidiary vector. The leading order expressions
turn out to be independent  of $\zeta$.~\footnote{The 
respective terms are
suppressed $\propto \sqrt{|\zeta^2/q^2|}$ and do therefore not contribute
at the twist--2 level but might be of relevance for the higher twist 
terms beginning with twist--3.} $e$ denotes the charge of the fermion 
field $\psi$, which is either a quark-- or an antiquark field.
$\hat{T}_{\mu\nu}(x)$,~Eq.~(\ref{Tmunu}), refers therefore to the 
contribution of
one of these fields for a single flavor. In the subsequent treatment
there is no essential structural difference considering quark or 
antiquark fields since we work in the massless approximation. The
expectation values of the bilocal quark and antiquark operators between 
nucleon states introduced below are of course different and do also
depend on the quark flavor. The complete Compton amplitude is obtained 
summing 
Eq.~(\ref{Tmunu}) over all quark and antiquark flavors contributing in
the kinematic domain considered.

The operators 
\begin{eqnarray}
\overline{\psi}\left(\frac{\tilde x}{2}\right)
\gamma^\mu \gamma^\lambda \gamma^\nu \psi
\left(-\frac{\tilde x}{2}\right)  \nonumber
\end{eqnarray}
are bilocal operators  on the light ray $ \tilde x $. These are
renormalized and time ordered operators~\cite{BGR}. One may rewrite the 
operator $\hat{T}_{\mu\nu}(x)$ in terms of a symmetric and an asymmetric
contribution by
\begin{eqnarray}
 \hat{T}_{\mu\nu}(x)  =
 - e^2 \frac{\tilde x^\lambda}{i \pi^2 (x^2-i\epsilon)^2}
 \left[S_{\alpha \mu\lambda \nu} 
 O^\alpha\left(\frac{\tilde x}{2}, -\frac{\tilde x}{2}\right)
+i \varepsilon_{\mu\lambda \nu \sigma} 
O_5^\alpha  \left(\frac{\tilde x}{2}, -\frac{\tilde x}{2}\right)
\right]~,
\end{eqnarray}
where
\begin{eqnarray}
S_{\alpha \mu\lambda \nu} = g_{\alpha\mu}g_{\lambda \nu}
                          + g_{\lambda\mu}g_{\alpha \nu}
                          - g_{\mu\nu}g_{\lambda \alpha}~.
\end{eqnarray}
The essential objects are the bilocal light--ray operators
\begin{eqnarray}
\label{oo}
O^{\alpha}\left(\frac{\tilde x}{2},-\frac{\tilde x}{2}\right)
&=&
\frac{i}{2}\left[\overline{\psi}\left(\frac{\tilde x}{2}\right)
\gamma^\alpha\psi\left(-\frac{\tilde x}{2}\right)
- \overline{\psi}\left(-\frac{\tilde x}{2}\right)
\gamma^\alpha\psi\left(\frac{\tilde x}{2}\right)\right]~,
\\
\label{oo5}
O^{\alpha}_5\left(\frac{\tilde x}{2},-\frac{\tilde x}{2}\right)
&=&
\frac{i}{2}
\left[\overline{\psi}\left(\frac{\tilde x}{2}\right)
\gamma_5\gamma^\alpha\psi\left(-\frac{\tilde x}{2}\right)
+ \overline{\psi}\left(-\frac{\tilde x}{2}\right)
\gamma_5\gamma^\alpha\psi\left(\frac{\tilde x}{2}\right)\right]~.
\end{eqnarray}
These operators do still contain higher twist contributions. In the 
present paper we are going to discuss the twist--2 contributions only
and have therefore to perform a twist decomposition.
The scalar twist--2 operators on the light--cone are  given by
\begin{eqnarray}
O\left(\frac{\tilde x}{2},-\frac{\tilde x}{2}\right)
&=&
\tilde{x}_{\alpha}\,
O^{\alpha}\left(\frac{\tilde x}{2},-\frac{\tilde x}{2}\right) \\
O_5\left(\frac{\tilde x}{2},-\frac{\tilde x}{2}\right)
&=&
\tilde{x}_{\alpha}\,
O^{\alpha}_5\left(\frac{\tilde x}{2},-\frac{\tilde x}{2}\right)~.
\end{eqnarray}
Their general definition off the light--cone can be found in 
Refs.~\cite{GLR,BGR}. The operators satisfy the condition
\begin{eqnarray}
\label{co1}
\Box O^{q,\;\rm traceless}_{(5)}(-\kappa x,\kappa x) = 0~.
\end{eqnarray}
Here $\kappa$ parametrizes   the position on the $x$--ray. The twist--2
vector operators can now be constructed referring to the 
scalar operator by
\begin{eqnarray}
\label{optw2}
O_{\sigma}^{q,\, \rm twist 2} (- \kappa \xx, \kappa \xx)
& = & \int_0^{1} d {\tau} \;
\partial_\sigma
\left.
O^q_{\rm traceless}(-\kappa\tau x, \kappa\tau x)
\right|_{x \rightarrow \xx}
\nonumber\\
& = & \int_0^{1} d {\tau}
 \left[ \partial_\sigma
+ \hbox{$\frac{1}{2}$}
(\ln \tau) x_\sigma \Box\right]
\left. O^q(-\kappa\tau x, \kappa\tau x)\right|_{x=\xx}~.
\end{eqnarray}
These operators satisfy the relations
\begin{eqnarray}
\label{co2}
\partial^\sigma O^{q,\;\rm traceless}_\sigma(-\kappa x,\kappa x)
= 0~,
\quad
\Box O^{q,\;\rm traceless}_\sigma(-\kappa x,\kappa x) = 0~.
\end{eqnarray}
\subsection{Operator Matrix Elements}
\renewcommand{\theequation}{\thesection.\arabic{equation}}
\label{sec-2.2}

\vspace{1mm}
\noindent
Up to now we have considered an expression for the product of the two 
electromagnetic currents. As the next step we form matrix elements of 
the respective operators. Let us first consider the matrix element of the
scalar operator. We use the kinematic decomposition into a {\sc Dirac}--
and a {\sc Pauli}--type contribution, the latter of which vanishes
in the case of forward scattering, $p_-, \eta \rightarrow 0$.
\begin{eqnarray}
e^2
\Biggl
\langle p_2, S_2 \left|O\left(\frac{x}{2}, - \frac{x}{2}\right)
\right |p_1, S_1 \Biggr \rangle
 &=&
i\,\overline{u}(p_2,S_2)\gamma x u(p_1,S_1) 
\int Dz e^{-ixp_z/2} f(z_1,z_2, p_i p_j x^2, p_i p_j,\mu^2_R) \nonumber\\
 &+&
i\,\overline{u}(p_2,S_2)x \sigma p_-  u(p_1,S_1) 
\int Dz e^{-ixp_z/2}
 g(z_1,z_2 , p_i p_j x^2, p_i p_j,\mu^2_R)   \nonumber\\
\end{eqnarray}
Here $\mu_R$ denotes the renormalization scale and
\begin{eqnarray}
\label{Dz}
Dz
&=& \frac{1}{2}
d z_1 d z_2 \theta(1-z_1) \theta(1+z_1) \theta(1-z_2) \theta(1+z_2) \\
&=&
d z_+ d z_-
\theta(1 + z_+ + z_-) \theta(1 + z_+ - z_-)
\theta(1 - z_+ + z_-) \theta(1 - z_+ - z_-)
\nonumber\\
p_z &=& p_+ z_+ +p_- z_- \nonumber\\ 
z_{\pm} &=& \frac{1}{2} ( z_2 \pm  z_1 ) \nonumber\\
\sigma_{\alpha\beta} &=& \frac{i}{2} \left[\gamma_\alpha \gamma_\beta
- \gamma_\beta \gamma_\alpha \right]~.
\end{eqnarray}
We consider all hadronic mass scales $p_i.p_j \approx 0$ as small 
compared to the large
invariants $p_{\pm}.q$ and $q^2$. Therefore the on--shell relations
\begin{eqnarray}
\gamma_\mu p_1^\mu~u(p_1,S_1) &=& 0 \nonumber\\
\overline{u}(p_2,S_2)~\gamma_\mu p_2^\mu &=& 0
\end{eqnarray}
hold. Under these assumptions we show that
the scalar matrix element satisfies the condition (\ref{co1}):
\begin{eqnarray}
\lefteqn{
\Box  e^2
\Biggl \langle p_2,S_2\left|O\left(\frac{x}{2},
- \frac{x}{2}\right)\right |p_1,S_1 \Biggr \rangle |_{x \rightarrow \tilde x}}
\\
&=&
i \Box \left [
 \overline{u}(p_2,S_2)\gamma x u(p_1,S_1) 
\int Dz e^{-i x p_z/2} f(z_1,z_2)  \right.
  \nonumber\\ & &
\left.
+
\overline{u}(p_2,S_2)x \sigma p_  u(p_1,S_1) 
\int Dz e^{-i x p_z/2}
 g(z_1,z_2 ) \right]|_{x \rightarrow \tilde x}
 \nonumber\\ 
&= &        
i\,\int Dz e^{-i\tilde x p(z)/2} f(z_1,z_2) 
\left[- \frac{i}{2} p_{z\mu} + 2\tilde{x}_{\mu}\left(-\frac{i}{2} p_z
\right)^2 \right]
\overline{u}(p_2,S_2)\gamma^\mu u(p_1,S_1)
\nonumber \\
&+& 
i\,\int Dz e^{-i\tilde x p(z)/2} f(z_1,z_2) 
\left[- \frac{i}{2} p_{z\mu} + 2\tilde{x}_{\mu}\left(-\frac{i}{2} p_z
\right)^2 \right]
\overline{u}(p_2,S_2)\sigma^{\mu\nu} p_{-\nu} u(p_1,S_1)          
\simeq 0~.
\nonumber
\end{eqnarray}
Analogously to the construction of the vector operator out of the scalar 
operator~(\ref{optw2}) one obtains the matrix element
\begin{eqnarray}
e^2
\Biggl \langle p_2,S_2 \left|O^{\mu}\left(\frac{{\tilde x}}{2},
- \frac{{\tilde x}}{2}\right)\right |p_1,S_1 \Biggr \rangle
&=&
i\,\int_0^1 d\lambda~\partial_x^\mu\Biggl\{
\overline{u}(p_2,S_2)\gamma x u(p_1,S_1) 
\int Dz e^{-i\lambda x p_z/2}
f(z_1,z_2) \nonumber\\
 & & ~~~~~+
\overline{u}(p_2,S_2)x \sigma p_- u(p_1,S_1)
\int Dz e^{-i \lambda  x p_z/2}
 g(z_1,z_2)  \Biggr\}|_{x \rightarrow \tilde x} \nonumber\\
&=& i\int Dz \partial_x^\mu \Biggl\{
\overline{u}(p_2,S_2)\gamma x u(p_1,S_1) 
 e^{-i x  p_z/2} \int_0^1 \frac{d\lambda}{\lambda^2}
f\left(\frac{z_1}{\lambda},\frac{z_2}{\lambda}\right) \nonumber\\
 & & ~~~~~+
\overline{u}(p_2,S_2)x \sigma p_-  u(p_1,S_1) 
 e^{-i x  p_z/2}\int_0^1 \frac{d\lambda}{\lambda^2}
g\left(\frac{z_1}{\lambda},\frac{z_2}{\lambda}\right)
 \Biggr\}|_{x \rightarrow \tilde x}.
 \nonumber\\
\end{eqnarray}
The expectation values of the operators $O^\mu$ and $Q^\mu_5$ are finally
expressed by
\begin{eqnarray}
\label{eqOM}
e^2
\lefteqn{
\Biggl \langle p_2,S_2 \left|O^{\mu}\left(\frac{{\tilde x}}{2},
- \frac{{\tilde x}}{2}\right)\right |p_1,S_1 \Biggr \rangle}  \\
&=& 
i\,\int Dz e^{-i\tilde x p_z/2} F(z_1,z_2)\left
[\overline{u}(p_2,S_2)\gamma^\mu  u(p_1,S_1) 
-\frac{i}{2} p_z^\mu  
\overline{u}(p_2,S_2)\gamma \tilde x  u(p_1,S_1) \right]
\nonumber
\\
&+&
i\int Dz e^{-i\tilde x p_z/2} G(z_1,z_2) \left
[\overline{u}(p_2,S_2)\sigma^{\mu\nu }{p_-}_\nu  u(p_1,S_1) 
-\frac{i}{2} p_z^\mu  \overline{u}(p_2,S_2)\sigma^{\alpha\beta}
\tilde x_\alpha {p_-}_\beta  u(p_1,S_1)\right]~.\nonumber
\end{eqnarray}
and
\begin{eqnarray}
\label{eqO5M}
\lefteqn{
e^2
\Biggl \langle p_2,S_2 \left|O^{\mu}_5\left(\frac{{\tilde x}}{2},
- \frac{{\tilde x}}{2}\right)\right |p_1,S_1 \Biggr \rangle}
\\
&=& 
\hspace*{-3mm}
i\,\int Dz e^{-i\tilde x p_z/2} F_5(z_1,z_2)\left
[\overline{u}(p_2,S_2)\gamma_5
\gamma^\mu  u(p_1,S_1) 
-\frac{i}{2} p_z^\mu  
\overline{u}(p_2,S_2)\gamma_5
\gamma \tilde x  u(p_1,S_1) \right]
\nonumber
\\
&+&
\hspace*{-3mm}
i\int Dz e^{-i\tilde x p_z/2} G_5(z_1,z_2) \left
[\overline{u}(p_2,S_2)\gamma_5
\sigma^{\mu\nu }{p_-}_\nu  u(p_1,S_1) 
-\frac{i}{2} p_z^\mu  \overline{u}(p_2,S_2)\gamma_5
\sigma^{\alpha\beta}
\tilde x_\alpha {p_-}_\beta  u(p_1,S_1)\right], 
\nonumber
\end{eqnarray}

respectively. The new functions $F,G \equiv H$ and $F_5,G_5 \equiv H_5$
are defined by
\begin{eqnarray}
H_{(5)}(z_1, z_2)=\int_0^1 \frac{d\lambda}{\lambda^2}
h_{(5)}\left(\frac{z_1}{\lambda},\frac{z_2}{\lambda}\right)~,
\end{eqnarray}
with $h_{(5)} \equiv f,g,f_5,g_5$. In the same manner as for the scalar 
operator it can be shown that the conditions (\ref{co2}) for the vector 
operator are satisfied by the matrix element. 

The functions $H$ and $H_5$ obey the relations
\begin{eqnarray}
\label{symrel1}
  H(z_1, z_2) &=& - H(-z_1,-z_2)\\
\label{symrel2}
H_5(z_1, z_2) &=& + H_5(-z_1,-z_2)~,
\end{eqnarray}
which follow  from Eqs.~(\ref{oo},\ref{oo5}) interchanging
$x$ and $-x$, in the limit in which contributions due to $\zeta$, 
Eq.~(\ref{xtil}), can be disregarded.

The contributions due to the above operators to the Compton amplitude 
read
\begin{eqnarray}
\label{COMPB}
T_{\mu\nu}(p_+,p_-,q)
 & = & i \int d^4x \,e^{iqx}\,
\langle p_2, S_2\,|T (J_{\mu}(x/2) J_{\nu}(-x/2))|\,p_1,
S_1 
\rangle~ \nonumber \\
 &=& \int d^4x \,e^{iqx}\,
\Biggl\{ -  \frac{\tilde x^\lambda}{i \pi^2 (x^2-i\varepsilon)^2}
 \Biggl[S_{\alpha \mu\lambda \nu}
\Biggl\langle p_2\Biggl
| O^\alpha\left(\frac{\tilde x}{2}, -\frac{\tilde x}{2}
\right)\Biggr
|p_1\Biggr\rangle
\nonumber \\
& &~~~~~~~~~~~~~~
+i \varepsilon_{\mu\lambda \nu \sigma}
\Biggl\langle p_2\Biggl
| O^\alpha_5\left(\frac{\tilde x}{2}, -\frac{\tilde x}{2}
\right)\Biggr
|p_1\Biggr\rangle   \Biggr]\Biggr\}~.
\end{eqnarray}
We rewrite Eq.~(\ref{COMPB}) referring to the distribution functions
$F_{(5)}, G_{(5)}$,
\begin{eqnarray}
\label{COMPB1}
\lefteqn{
T_{\mu\nu}(p_+,p_-,q)
  =  - \int d^4x \,e^{i(q- p_z/2)x}\,
      \frac{\tilde x^\lambda}{  \pi^2 (x^2-i\varepsilon)^2} }  \\
 & & \times \Biggl\{
     S_{\alpha \mu\lambda \nu} \Biggl\{
\int Dz \left[
\overline{u}(p_2,S_2)\gamma^\alpha  u(p_1,S_1)
- \frac{i}{2} p_z^\mu  
\overline{u}(p_2,S_2)\gamma \tilde x  u(p_1,S_1) \right] F(z_+,z_-)
\nonumber\\
& & +
\int Dz  
\left[\overline{u}(p_2,S_2)\sigma^{\alpha\beta}{p_-}_\beta  u(p_1,S_1)
-\frac{i}{2} p_z^\alpha  \overline{u}(p_2,S_2)\sigma^{\beta\gamma}
\tilde x_\beta {p_-}_\gamma  u(p_1,S_1) \right]
G(z_+,z_-) 
\Biggr\}  \nonumber \\
& & +i \varepsilon_{\mu\lambda \nu \sigma} \Biggl\{
\int Dz 
\left[\overline{u}(p_2,S_2)\gamma_5\gamma^\sigma  u(p_1,S_1)
-\frac{i}{2}  p_z^\sigma   \overline{u}(p_2,S_2)\gamma_5
\gamma \tilde x  u(p_1,S_1) \right] F_5(z_+,z_-)
\nonumber\\
& & +
\int Dz  
\left[\overline{u}(p_2,S_2)\gamma_5\sigma^{\sigma\alpha}{p_-}_\alpha 
u(p_1,S_1)
- \frac{i}{2} p_z^\sigma   \overline{u}(p_2,S_2)\gamma_5
\sigma^{\alpha\beta }
\tilde x_\alpha {p_-}_\beta  u(p_1,S_1)\right]
G_5(z_+,z_-) \Biggr\} \Biggr \}~. \nonumber
\end{eqnarray}
Finally the Fourier transform to momentum space is performed,
\begin{eqnarray}
\label{comp4}
T_{\mu\nu}(q,p_+,p_-)
 &=&  -2
 \int Dz
 \frac{1}{Q^2 + i \varepsilon} \Biggl\{
\overline{u}(p_2,S_2)\Gamma^F_{\mu\nu}(q,p_+,p_-) u(p_1,S_1)~F(z_+,z_-)
\nonumber\\ &&~~~~~~~~~~~~~~~~~~~~
+ \overline{u}(p_2,S_2)\Gamma^{F5}_{\mu\nu}(q,p_+,p_-) 
u(p_1,S_1)~F_5(z_+,z_-) \nonumber\\   &&~~~~~~~~~~~~~~~~~~~~
+\overline{u}(p_2,S_2)\Gamma^G_{\mu\nu}(q,p_+,p_-) u(p_1,S_1)~G(z_+,z_-)
\nonumber\\ &&~~~~~~~~~~~~~~~~~~~~
+\overline{u}(p_2,S_2)\Gamma^{G5}_{\mu\nu}(q,p_+,p_-) 
u(p_1,S_1)~G_5(z_+,z_-) \Biggr\}~.
\label{comp1}
\end{eqnarray}
The matrices $\Gamma_{\mu\nu}^O$  are given by
\begin{eqnarray}
\Gamma^F_{\mu\nu}(q,p_z) &=& ~~\Bigl [
Q_{\mu} \gamma_{\nu}  +
Q_{\nu} \gamma_{\mu}  -
g_{\mu\nu} \gamma_{\alpha} Q^{\alpha}  \Bigr]
\nonumber\\ &&
- \frac{1}{2}
\Bigl [
p_{z \mu} \gamma_{\nu} +
p_{z \nu} \gamma_{\mu} - g_{\mu\nu} \gamma_{\alpha} p_z^{\alpha} \Bigr]
\nonumber\\ && +
\frac{1}{Q^2+i\varepsilon} \gamma_{\alpha} Q^{\alpha} 
\Bigl [Q_{\nu} p_{z \mu} + Q_{\mu} p_{z \nu} - g_{\mu\nu} Q.p_z \Bigr]
\nonumber\\
                &\simeq&         ~~\Bigl [
q_{\mu} \gamma_{\nu}  +
q_{\nu} \gamma_{\mu}  -
g_{\mu\nu} \gamma_{\alpha} q^{\alpha}  \Bigr]
- \left[p_{z\mu} \gamma_\nu + p_{z\nu} \gamma_\mu\right]
\nonumber\\ &&
+ \frac{1}{Q^2+i\varepsilon} \gamma_{\alpha} q^{\alpha}
\Bigl [
     -p_{z \nu} p_{z \mu} +
q_{\nu} p_{z \mu} + q_{\mu} p_{z \nu} - g_{\mu\nu} q.p_z
\Bigr]
\\                               
\Gamma^{F5}_{\mu\nu}(q,p_z) &=& i~\gamma_5
\varepsilon_{\mu\nu\lambda\sigma}
\left[Q^{\lambda} \gamma^{\sigma}
- \frac{1}{2} p_{z}^{\sigma} \gamma^{\lambda}
+ \frac{1}{Q^2 + i \varepsilon} Q^{\lambda}
p_{z}^{\sigma} \gamma_{\alpha} Q^{\alpha}
\right] \nonumber\\
                                &\simeq&
i~\gamma_5
\varepsilon_{\mu\nu\lambda\sigma}
\left[q^{\lambda} \gamma^{\sigma}
+ \frac{1}{Q^2 + i \varepsilon} q^{\lambda}
p_{z}^{\sigma} \gamma_{\alpha} q^{\alpha}
\right]     
\\
\Gamma^{G}_{\mu\nu}(q,p_z) &=& ~~\Bigl [
Q_{\mu} \sigma_{\nu\alpha} p_-^{\alpha} +
Q_{\nu} \sigma_{\mu\alpha} p_-^{\alpha} -
g_{\mu\nu} \sigma_{\alpha\beta} p_-^{\beta} Q^{\alpha}  \Bigr]
\nonumber\\ &&
- \frac{1}{2}
\Bigl [
p_{z \mu} \sigma_{\nu\alpha} p_-^{\alpha} +
p_{z \nu} \sigma_{\mu\alpha} p_-^{\alpha} -
g_{\mu\nu} \sigma_{\beta\alpha} p_-^{\alpha} p_z^{\beta}  \Bigr]
\nonumber\\ && +
\frac{1}{Q^2+i\varepsilon} \sigma_{\beta\alpha} p_-^{\alpha} Q^{\beta}
\Bigl [Q_{\nu} p_{z \mu} + Q_{\mu} p_{z \nu} - g_{\mu\nu} Q.p_z \Bigr]
\nonumber\\
                               &\simeq&~~~\Bigl [
q_{\mu} \sigma_{\nu\alpha} p_-^{\alpha} +
q_{\nu} \sigma_{\mu\alpha} p_-^{\alpha} -
g_{\mu\nu} \sigma_{\beta\alpha} p_-^{\alpha} q^{\beta}  \Bigr]
-\Bigl [
p_{z \mu} \sigma_{\nu\alpha} p_-^{\alpha} +
p_{z \nu} \sigma_{\mu\alpha} p_-^{\alpha} \Bigr]
\nonumber\\ & & +
\frac{1}{Q^2+i\varepsilon} \sigma_{\beta\alpha} p_-^{\alpha} q^{\beta}
\Bigl [-p_{z \mu} p_{z \nu} +
q_{\nu} p_{z \mu} + q_{\mu} p_{z \nu} - g_{\mu\nu} q.p_z \Bigr]
\nonumber\\
\\
\Gamma^{G5}_{\mu\nu}(q,p_z) &=& i~\gamma_5
\varepsilon_{\mu\nu\lambda\sigma}
\left[Q^{\lambda} \sigma^{\sigma\alpha} p_{-\alpha}
- \frac{1}{2} p_{z}^{\sigma} \sigma^{\lambda\alpha} p_{-\alpha}
+ \frac{1}{Q^2 + i \varepsilon} Q^{\lambda}
p_{z}^{\sigma} \sigma^{\alpha\beta} Q_{\alpha} p_{-\beta}
\right] \nonumber\\
                                &\simeq&
i~\gamma_5
\varepsilon_{\mu\nu\lambda\sigma}
\left[q^{\lambda} \sigma^{\sigma\alpha} p_{-\alpha}
+ \frac{1}{Q^2 + i \varepsilon} q^{\lambda}
p_{z}^{\sigma} \sigma^{\alpha\beta} q_{\alpha} p_{-\beta}
\right]~,
\end{eqnarray}
where
\begin{eqnarray}
Q=q-\frac{p_z}{2}~.
\end{eqnarray}
The equivalence--sign denotes that the other contributions vanish between
the bi--spinor states according to the assumption that 
$p_i.p_j \approx 0$.
\subsection{Lorentz Structure}
\renewcommand{\theequation}{\thesection.\arabic{equation}}
\label{sec-2.3}

\vspace{1mm}
\noindent
Let us now discuss the Lorentz structure of the Compton amplitude in 
more detail. The matrices $\Gamma^{O}_{\mu\nu}/(Q^2 + i \varepsilon)$ 
consist out of two parts which are $\propto 1/Q^2$ and
$\propto 1/Q^4$, respectively. The numerators of the former terms
depend on $q$ only, whereas the second terms contain the vectors $q$ and
$p_z = z_+ p_+ + z_- p_-$. The $(z_+,z_-)$--dependence may be factored 
out of the quantities
\begin{eqnarray}
\overline{u}(p_2,S_2)\gamma_{\mu_1} \ldots \gamma_{\mu_k} (\gamma_5)
u(p_1,S_1)
\end{eqnarray}
and the form factors related to the contributions $\propto 1/Q^2$ and
$\propto 1/Q^4$ are of different structure, graded by the 
$z_{\pm}$ dependence in the numerators. Due to the different
numerator structure individual variations in $q$, $p_+$ and $p_-$ may
allow to disentangle the different contributions experimentally. 
The Compton amplitude is given by
\begin{eqnarray}
\label{comp5}
T_{\mu\nu}(q,p_+,p_-)
 &=& -2
 \overline{u}(p_2,S_2) \left[
 \Gamma_{\mu\nu}^{F}(q,p_+,p_-)+
 \Gamma_{\mu\nu}^{F5}(q,p_+,p_-)  \right.
 \nonumber\\  
 & &~~~~~~~~~~~~\left.
 +\Gamma_{\mu\nu}^{G}(q,p_+,p_-)+
 \Gamma_{\mu\nu}^{G5}(q,p_+,p_-)\right]
 u(p_1,S_1)~,
\label{comp1a}
\end{eqnarray}
with
\begin{eqnarray}
\label{galo1}
\Gamma^F_{\mu\nu}(q,p_+,p_-) &=& 
\Bigl [
q_{\mu} \gamma_{\nu}  +
q_{\nu} \gamma_{\mu}  -
g_{\mu\nu} \gamma_{\alpha} q^{\alpha}  
\Bigr] F_1(\xi,\eta) \nonumber\\ & &
- \gamma_\mu F_{1,\nu}(\xi,\eta)
- \gamma_\nu F_{1,\mu}(\xi,\eta)
+ \gamma_{\alpha} q^{\alpha} F_{2,\mu\nu}(\xi,\eta)
\\                               
\Gamma^{F5}_{\mu\nu}(q,p_+,p_-) &=& 
i~\gamma_5
\varepsilon_{\mu\nu\lambda\sigma}
\left[q^{\lambda} \gamma^{\sigma} F^5_1(\xi,\eta)
+ q^{\lambda}  \gamma_{\alpha} q^{\alpha} F_2^{\sigma,5}(\xi,\eta)
\right]\\
\Gamma^G_{\mu\nu}(q,p_+,p_-) &=&
\Bigl [
q_{\mu} \sigma_{\nu\alpha} p_-^{\alpha} +
q_{\nu} \sigma_{\mu\alpha} p_-^{\alpha}
-
g_{\mu\nu} \sigma_{\beta\alpha} p_-^{\alpha} q^{\beta} 
\Bigr] 
G_1(\xi,\eta) \nonumber\\ & &
- \sigma_{\mu\alpha} p_-^\alpha G_{1,\nu}(\xi,\eta)
- \sigma_{\nu\alpha} p_-^\alpha G_{1,\mu}(\xi,\eta)
p_-^\alpha
+ \sigma_{\beta\alpha} p_-^{\alpha} q^{\beta} G_{2,\mu\nu}(\xi,\eta)
\\                               
\label{galo2}
\Gamma^{G5}_{\mu\nu}(q,p_+,p_-) &=&
i~\gamma_5
\varepsilon_{\mu\nu\lambda\sigma}
\left[q^{\lambda} \sigma^{\sigma\alpha} p_{-\alpha} G_1^5(\xi,\eta)+
 q^{\lambda}  \sigma^{\alpha\beta} q_{\alpha} p_{-\beta} 
 G_2^{\sigma,5}(\xi,\eta)
\right]~.
\end{eqnarray}
The different amplitude functions are given by
\begin{eqnarray}
H_1(\xi,\eta) &=& \int Dz \frac{1}{Q^2+i\varepsilon} H(z_+,z_-) \\
\label{eqH1}
H_k^{\sigma}(\xi,\eta) &=& \int Dz \frac{p_+^{\sigma} z_+
+ p_-^{\sigma} z_-}{(Q^2+i\varepsilon)^k} H(z_+,z_-) =
                           \int Dz \frac{p_+^{\sigma} t + \pi_{\sigma}
z_-}             
{(Q^2+i\varepsilon)^k} H(z_+,z_-)
\\
\label{eqH2}
H_{2,\mu\nu}(\xi,\eta) 
&=& \int Dz \frac{1}{(Q^2+i\varepsilon)^2}
\left[
-p_{z\mu}p_{z\nu}
    + q_{\nu} p_{z\mu} +q_{\mu} p_{z\nu} -g_{\mu\nu} q.p_z\right]
H(z_+,z_-)          \\
&=& \int Dz \frac{1}{(Q^2+i\varepsilon)^2}
\left[
-p_{+\mu}p_{+\nu} t^2
    + \left(
q_{\nu} p_{+\mu} +q_{\mu} p_{+\nu} \right)
t -g_{\mu\nu} q.p_z
\right. \nonumber\\ & &~~~~~~~~~~~~~~~~~~~~~~~ \left.
-\pi_{\mu}\pi_{\nu} z_-^2
    + \left(q_{\nu} \pi_{\mu} +q_{\mu} \pi_{\nu}\right) z_-
    + \left(p_{+\nu} \pi_{\mu} +p_{+\mu} \pi_{\nu}\right) t z_-
\right] \nonumber\\ & & 
~~~~~~~~~~~~~~~~~~~~~~~~~~~~~~~~~~~~~~~~~~~~~~~~~~~~~~~~~~~~~~~~~~~~~~
~~~~~~~~~~~~~\times
H(z_+,z_-)~,  \nonumber
\end{eqnarray}
where
\begin{eqnarray}
t &=& z_+ + \eta z_-\\
\pi_\sigma &=& p_{-\sigma} - \eta p_{+\sigma}
\end{eqnarray}
and $q.p_z = q.p_+ t$.
Here $H$ denotes either $F$ or $G$. 
The amplitude functions  in Eqs.~(\ref{galo1}--\ref{galo2}) depend on the 
scaling variables $\xi$ and $\eta$. In the limit of forward scattering  
($\eta \rightarrow 0, \xi \rightarrow x_B$) the structure functions
$F_{1,2}(x_B)$ and $g_{1,2}(x_B)$ occur as the limit of the amplitude
functions of respective helicity projections from the {\sc Dirac}--type 
terms. The relations between these amplitude
functions are derived in
Section~\ref{sec-6}.
\section{Kinematic Relations}
\renewcommand{\theequation}{\thesection.\arabic{equation}}
\setcounter{equation}{0}
\label{sec-3}

\vspace{1mm}
\noindent
For convenience we construct the helicity basis for the photons 
$\gamma_1^*$ and $\gamma_2^{(*)}$ in the Breit--frame
\begin{eqnarray}
p_+ &=& p_1 + p_2 = (2 E_p;~\vec{0}) \nonumber\\
-p_-&=& p_1 - p_2 = (0;~2 \vec{p})  = (0;~0,0,2p_3)
\nonumber\\
q &=&  \hbox{\large $\frac{1}{2}$} \left(q_1 + q_2\right) 
= (q_0;~q_1,0,q_3)~.
\end{eqnarray}
The matrix elements 
\begin{equation}
\T_{kl} = \epsi_{2,k}^\mu T_{\mu\nu} \epsi_{1,l}^{\nu},~~k,l~\epsilon~
\{0,1,2,3\}
\end{equation}
are independent of the reference frame.

In the generalized Bjorken region we classify the various contributions
to $\T_{kl}$ by their $\nu$--dependence, where the leading terms are
kept respectively. There the kinematic invariants are given by
\begin{eqnarray}
\label{eqq1}
q_1.q_1 &=& - \nu (\xi-\eta)     \\
\label{eqq2}
q_2.q_2 &=& - \nu (\xi+\eta)     \\
q.p_+ &=& \nu  \\
q.p_- &=& \eta \nu  \\
q.q   &=& -\xi \nu  \\
q.p_z &=& q^2-Q^2 = (z_+  + z_- \eta ) \nu \equiv t~\nu  \\
p_+^2 &\approx& p_-^2 \approx p_+ p_- \approx 0~.
\end{eqnarray}

To define the helicity basis we introduce the two reference vectors
\begin{eqnarray}
n_0 &=& (1;~0,0,0)\\
n_2 &=& (0;~0,1,0)~.
\end{eqnarray}
The polarization vectors of the photons $\gamma_1^*$ and $\gamma_2^*$
are then given by
\begin{eqnarray}
\label{eqep1}
\varepsilon_{0 \mu}^{(1)} &=& \frac{q_{1\mu}}{\sqrt{|q_1^2|}}
= \frac{q_{1\mu}}{\nu^{1/2}} \frac{1}{\sqrt{|\xi-\eta|}}
\nonumber\\
\varepsilon_{0 \mu}^{(2)} &=& \frac{q_{2\mu}}{\sqrt{|q_2^2|}}
= \frac{q_{2\mu}}{\nu^{1/2}} \frac{1}{\sqrt{|\xi+\eta|}}
\\
\varepsilon_{1 \mu}^{(i)} &=& n_{2\mu}\\
\varepsilon_{2 \mu}^{(i)} &=& \frac{1}{N_{2i}}
\varepsilon_{\mu \alpha \beta \gamma} n_0^{\alpha} n_2^{\beta}
q_i^{\gamma}  \\
\label{eqep2}
\varepsilon_{3 \mu}^{(i)} &=& \frac{1}{N_{3i}}
\left[q_{i\mu} q_i.n_0 - n_{0\mu} q_i.q_i \right]~,
\end{eqnarray}
with $i = 1,2$ and
\begin{eqnarray}
N_{21} &=& \frac{\nu}{\mu}
\sqrt{\left|1 + \frac{\mu^2}{\nu}(\xi-\eta)\right|}
\nonumber\\
N_{22} &=& \frac{\nu}{\mu} 
\sqrt{\left|1 + \frac{\mu^2}{\nu}(\xi+\eta)\right|}
\nonumber\\
N_{31} &=& \frac{\nu^{3/2}}{\mu} \sqrt{|\xi - \eta|}
\sqrt{\left|1 + \frac{\mu^2}{\nu}(\xi-\eta)\right|}
\nonumber\\
N_{32} &=& \frac{\nu^{3/2}}{\mu} \sqrt{|\xi + \eta|}
\sqrt{\left|1 + \frac{\mu^2}{\nu}(\xi+\eta)\right|}~,
\end{eqnarray}
with $\mu^2 = p_+^2$.

With respect to the vector $q_2$, Eq.~(\ref{eqq2}),
the above notation applies
for space- or time-like vectors, for which
$|\xi \pm \eta| \neq 0$ and
all the above terms are regular.
The polarization vectors obey
\begin{eqnarray}
\varepsilon_{k \mu}^{(i)}.
\varepsilon_{l}^{(i) \mu} =   s_k \delta_{kl}~,
\end{eqnarray}
with $s_k=-1$ for $k=0,1,2$ and $s_k=+1$ for $k=3$ if both vectors are
space-like and analogous  relations if $q_2$ is time-like.
$\varepsilon_{1,2\mu}^{(i)}$ are the transversal and
$\varepsilon_{3\mu}^{(i)}$ is the longitudinal polarization vector of the
photon $\gamma_i^*$.

As will be shown below the hadronic mass scale $\mu^2 = p_+^2$ occurring
as a normalization parameter cancels in the helicity projections
of the Compton amplitude in leading order in $1/\nu$, 
cf. Section~\ref{sec-5}.

For the limit to the forward case and the understanding of the results
in the limit $\mu^2 \ll \nu$ it is useful to rewrite the  polarization
vectors (\ref{eqep1}--\ref{eqep2}) as
\begin{eqnarray}
\varepsilon_{0 \rho}^{(1(2))} &=&  \frac{1}{\sqrt{|\xi|}}
\left[\frac{q_{\rho}}{\nu^{1/2}}
\pm \frac{p_{-\rho}}{2 \nu^{1/2}}\right]
 \frac{1}{\sqrt{|1\mp\eta/\xi|}}
\\
\varepsilon_{1 \rho}^{(1(2))} &=& n_{2\rho}\\
\varepsilon_{2 \rho}^{(1(2))} &=& \frac{\mu}{\nu} \left|1 - \frac{\mu^2}
{2\nu} (\xi \mp \eta)\right| \varepsilon_{\rho\alpha\beta\gamma}
n_0^\alpha n_2^\beta \left(q^\gamma \pm \frac{1}{2} p_-^\gamma\right)\\
\varepsilon_{3 \rho}^{(1(2))} &=& 
\frac{1}{\nu^{1/2}} \frac{1}
{\sqrt{|\xi|}} \left|1 - \frac{\mu^2}{2\nu} (\xi \mp \eta)\right|
\left[q_\rho \pm \frac{1}{2} p_{-\rho}
+ \mu n_{0\rho} (\xi \mp \eta)\right]
\frac{1}{\sqrt{|1\mp\eta/\xi|}}~.
\end{eqnarray}
In the limit $p_{-\rho}, \eta \rightarrow 0$ the polarization vectors of 
the first and the second photon become identical. On the other hand,
in the limit that terms of
$O(\mu^2)$ can be neglected against terms of $O(\nu)$ the polarization
vectors $\varepsilon_{3\rho}^{(i)}$ and $\varepsilon_{0\rho}^{(i)}$
become identical for $i = 1,2$. The latter aspect induces that current
conservation for the twist--2 contributions, cf.~section~4, enforces
that helicity projections in the direction of the longitudinal 
polarization
vectors vanish in the generalized Bjorken region in lowest order QCD.

If $q_2$ is light--like the respective set of polarization vectors
reads
\begin{eqnarray}
\label{eqep4a}
\varepsilon_{0 \mu}^{(2)} &=& 
\frac{1}{\sqrt{2} q_0^{(2)}} q_{\mu} =
\frac{1}{\sqrt{2} q_0^{(2)}} (q_0,~\vec{q}_2)\\
\varepsilon_{1 \mu}^{(2)} &=& n_{2\mu}\\
\label{eqep4b}
\varepsilon_{2 \mu}^{(2)} &=& \frac{1}{q_0^{(2)}} 
\varepsilon_{\mu\alpha\beta\gamma} n_0^\alpha n_2^\beta q_2^\gamma~,
\end{eqnarray}
with
\begin{eqnarray}
\label{eqep5}
q_0^{(2)} = \frac{\nu}{\mu}~.
\end{eqnarray}
A fourth linearly independent vector associated to the above set is
\begin{eqnarray}
\label{eqep6}
\widetilde{\varepsilon}_{0 \mu}^{(2)} &=&
\frac{1}{\sqrt{2} q_0^{(2)}} (q_0,~-\vec{q}_2)
\end{eqnarray}
with
\begin{eqnarray}
\label{eqep7}
\widetilde{\varepsilon}_{0 \mu}^{(2)} + 
\varepsilon_{0 \mu}^{(2)} = \sqrt{2} n_{0\mu}~.
\end{eqnarray}
The vectors $\widetilde{\varepsilon}_{0 \mu}^{(2)},
\varepsilon_{0 \mu}^{(2)}, \varepsilon_{1 \mu}^{(2)}$ and
$\varepsilon_{2 \mu}^{(2)}$ span the Minkowski space.
For later use we also note that the vectors
\begin{eqnarray}
p_+^\mu~,p_-^\mu~,~q^\mu~~~~{\rm and}~~~~n_2^\mu \nonumber
\end{eqnarray}
span the Minkowski space for the general case of non--forward
scattering, excluding special cases as forward scattering $p_- = 0$ and
vacuum--meson transition $p_+=p_-$.
\section{Current Conservation}
\renewcommand{\theequation}{\thesection.\arabic{equation}}
\setcounter{equation}{0}
\label{sec-4}

\vspace{1mm}
\noindent
The conservation of the electromagnetic current
\begin{eqnarray}
\partial_\mu^x J^\mu(x) = 0
\end{eqnarray}
implies for the Compton amplitude
\begin{eqnarray}
T_{\mu\nu}(p_+,p_-,q) &=&
i\,\int d^4 e^{-iq_2 x} \langle p_2,S_2|RT(J_{\mu}(0)J_{\nu}(x)|p_1,S_1
\rangle \nonumber\\ &=&
i\,\int d^4 e^{-iq_1 x} \langle p_2,S_2|RT(J_{\mu}(-x)J_{\nu}(0)|p_1,S_1
\rangle 
\end{eqnarray}
the relations
\begin{eqnarray}
q_2^{\mu} T_{\mu\nu} = T_{\mu\nu} q^{\nu}_1 = 0~.
\end{eqnarray}
Expanding the Compton amplitude according to the (non--local) operator 
product expansion for deep inelastic non--forward scattering in the 
generalized Bjorken region the terms obtained forming the matrix elements
(\ref{COMPB1}) are not necessarily yet the twist--2 contributions only. 
The explicit calculation shows, that still terms of the order 
$(\mu^2/\nu)^{1/2+k},~k \geq 0$ are contained. These terms are of higher
twist and vanish for $\nu \rightarrow \infty$ if compared to the leading
twist--2 contributions. To prove the current conservation for the
twist--2 contributions to the Compton amplitude these terms have to be
dealt with in common with the respective higher twist contributions
resulting from the operator matrix elements of the higher twist
operators.

We are firstly considering the helicity projections of the Compton
amplitude onto the states $\varepsilon_{0\mu}^{(2)}$ and
$\varepsilon_{0\nu}^{(1)}$, respectively, if $q_2$ is either space- 
or time-like.
 The remaining index is
contracted by all the four helicity vectors. As each of the sets
of four helicity vectors corresponding to the virtual photons 
$\gamma_{1,2}^*$ spans the complete Minkowski--space, vanishing
of the
projections in all components proves the current conservation for
the twist--2 contributions. We list the individual projections for the
contributions to the different distribution functions $F,F_5,G$ and $G_5$
separately.
\begin{eqnarray}
\label{eqgau1}
\T^F_{00} &=& -
\frac{2}{\nu} \overline{u}(p_2,S_2) \gamma_{\mu} q^{\mu} u(p_1,S_1)
\frac{1}{\sqrt{|\xi^2-\eta^2|}}
\int Dz~ F(z_+,z_-)\\
\T^F_{01},
\T^F_{10},
\T^F_{02},
\T^F_{20}
&=& O\left(\frac{1}{\sqrt{\nu}}\right)\\
\T^F_{03} &=& -
 \frac{2}{\nu} \overline{u}(p_2,S_2) \gamma_{\mu} q^{\mu} u(p_1,S_1)
\frac{1}{|\xi-\eta| \sqrt{|\xi^2-\eta^2|}}
\int Dz~ F(z_+,z_-)\\
\T^F_{30} &=&  -
 \frac{2}{\nu} \overline{u}(p_2,S_2) \gamma_{\mu} q^{\mu} u(p_1,S_1)
\frac{1}{|\xi+\eta| \sqrt{|\xi^2-\eta^2|}}
\int Dz~ F(z_+,z_-)\\
\T^{F5}_{00} &=&  0\\
\T^{F5}_{01},
\T^{F5}_{10},
\T^{F5}_{02},
\T^{F5}_{20} &=& O\left(\frac{1}{\sqrt{\nu}}\right)\\
\T^{F5}_{03},
\T^{F5}_{30} &=& O\left(\frac{1}{\nu}\right)\\
\T^G_{00} &=& -
\frac{2}{\nu} \overline{u}(p_2,S_2) \gamma_{\mu} q^{\mu} u(p_1,S_1)
\frac{1}{\sqrt{|\xi^2-\eta^2|}}
\int Dz~ G(z_+,z_-)
\\
\T^G_{01},
\T^G_{10},
\T^G_{02},
\T^G_{20}
&=& O\left(\frac{1}{\sqrt{\nu}}\right)\\
\T^G_{03} &=&  -
 \frac{2}{\nu} \overline{u}(p_2,S_2) \sigma_{\beta\alpha}  p_-^{\alpha}
q^{\beta} u(p_1,S_1)
\frac{1}{|\xi-\eta| \sqrt{|\xi^2-\eta^2|}}
\int Dz~ G(z_+,z_-)\nonumber\\
\\
\T^G_{30} &=& -
 \frac{2}{\nu} \overline{u}(p_2,S_2) \sigma_{\beta\alpha}  p_-^{\alpha}
q^{\beta} u(p_1,S_1)
\frac{1}{|\xi+\eta| \sqrt{|\xi^2-\eta^2|}}
\int Dz~ G(z_+,z_-)\nonumber\\ \\
\T^{G5}_{00} &=&  0\\
\T^{G5}_{01},
\T^{G5}_{10},
\T^{G5}_{02},
\T^{G5}_{20} &=& O\left(\frac{1}{\sqrt{\nu}}\right)\\
\label{eqgau2}
\T^{G5}_{03},
\T^{G5}_{30} &=& O\left(\frac{1}{\nu}\right)~.
\end{eqnarray}
Note that
\begin{eqnarray}
\overline{u}(p_2,S_2) \gamma_{\mu} q^{\mu} u(p_1,S_1) &\propto& \nu
\nonumber\\
\varepsilon_{\alpha,\beta,\gamma,\delta}~p_{\pm}^{\gamma} q^{\delta}
&\propto& \nu~,
\end{eqnarray}
etc. The contributions to ${\sf T}_{kl}^{H(5)}$ of $O(1/\sqrt{\nu})$ or
$O(1/\nu)$ are of higher twist since the expectation values of the
twist--2 operators are multiplied by a further mass ratio, cf.~\cite{BN}.
Since the integrals
\begin{eqnarray}
\label{eqint}
\int~Dz~H(z_+,z_-) = \int_{-1}^{+1} d z_+ \int_{-1+|z_+|}^{+1-|z_+|}
H(z_+,z_-) = 0
\end{eqnarray}
for $H = F,~G$ vanish, all projections Eqs.~(\ref{eqgau1}--\ref{eqgau2})
vanish in the Bjorken limit proving current conservation for the
twist--2 contributions in the generalized Bjorken region.

In the case of forward scattering the evaluation of
the current--current operator in terms of the twist--2 operators 
current conservation is easily obtained, cf. Ref.~\cite{BT}. In the limit
$p_-, \eta \rightarrow 0$ the above expressions 
(\ref{eqgau1}--\ref{eqgau2}) vanish for all values of $\nu$ by virtue
of Eq.~(\ref{eqint}).  Unlike the forward case which is characterized
by only two kinematic vectors $q$ and $p$, the presence of the vector
$p_-$ in the non--forward case allows for twist--3 terms also for 
spin--averaged matrix elements. These are of $O(1/\sqrt{\nu})$ relative 
to the twist--2 contributions.

The above conclusions hold analogously if $q_2$ is a light--like vector.
Here the polarization vectors 
Eqs.~(\ref{eqep4a}--\ref{eqep4b}, \ref{eqep6}) have to be used.
\section{The Helicity Projections of the Compton Amplitude}
\renewcommand{\theequation}{\thesection.\arabic{equation}}
\setcounter{equation}{0}
\label{sec-5}

\vspace{1mm}
\noindent
In the following we calculate the helicity projections of the twist--2 
contributions to the  Compton amplitude (\ref{comp1a}) representing the 
matrices $\Gamma_{\mu\nu}^O$ (\ref{galo1}--\ref{galo2}) in the 
generalized Bjorken region. The unpolarized {\sc Dirac}--type terms 
are\footnote{Here and in the following we will use the terms
{\sf unpolarized} and {\sf polarized} for the symmetric and anti-symmetric
contributions to the Compton amplitude, respectively. In the 
spin-averaged
case the anti-symmetric terms cancel.}
~:
\begin{eqnarray}
\label{TF11L}
\T^F_{11} &=&
2 \overline{u}(p_2,S_2) \gamma_{\mu} q^{\mu} u(p_1,S_1)
  \left[F_1(\xi,\eta) + \varepsilon_{1}^{(2)\mu} \varepsilon_{1}^{(1)\nu}
  F_{2,\mu\nu}(\xi,\eta)\right]\\
\label{TF22L}
\T^F_{22} &=&
2 \overline{u}(p_2,S_2) \gamma_{\mu} q^{\mu} u(p_1,S_1)
  \left[F_1(\xi,\eta) 
  + \varepsilon_{2}^{(2)\mu} \varepsilon_{2}^{(1)\nu}
  F_{2,\mu\nu}(\xi,\eta)\right]\\
\T^{F}_{kl} &\propto& \left(\frac{1}{\nu}\right)
^{1/2+n}~~~~~~{\rm for~the~other~projections}~~{k,l~\epsilon~\{1,2,3\}}~~
{\rm and}~~~{n \geq 0}~,
\end{eqnarray}
with
\begin{eqnarray}
\label{eqF2H}
\varepsilon_{1}^{(2)\mu} \varepsilon_{1}^{(1)\nu}
F_{2\mu\nu}(\xi,\eta) = 
\varepsilon_{2}^{(2)\mu} \varepsilon_{2}^{(1)\nu}
F_{2\mu\nu}(\xi,\eta) = 
\int Dz \frac{q.p_z}{(Q^2+i\varepsilon)^2} F(z_+,z_-)~,
\end{eqnarray}
leading to
\begin{eqnarray}
\T_{11}^F = \T_{22}^F~.
\end{eqnarray}
Correspondingly the polarized {\sc Dirac}--type terms yield~:
\begin{eqnarray}
\label{eqpol}
\T_{12}^{F5} = -\T_{21}^{F5}
\end{eqnarray}
\begin{eqnarray}
\T^{F5}_{kl} &\propto& \left(\frac{1}{\nu}\right)
^{1/2+n}~~~~~~{\rm for~the~other~projections}~~{k,l~\epsilon~\{1,2,3\}}~~
{\rm and}~~~{n \geq 0}~.
\end{eqnarray}
We define the vector
\begin{eqnarray}
S_{21}^{\sigma} := 
- \frac{1}{2} \overline{u}(p_2,S_2) \gamma_5 \gamma^{\sigma} u(p_1,S_1)~.
\end{eqnarray}
In the case of forward scattering, $(p_2,S_2) \rightarrow (p_1,S_1)$,
$S_{21}^{\sigma}$ denotes the nucleon spin vector $S^{\sigma}$.
$\T^{F5}_{12}$ is given by
\begin{eqnarray}
\label{EQF5}
\T^{F5}_{12} &=& i~\varepsilon^{\mu\lambda\nu\sigma}
\varepsilon_{1\mu}^{(2)} \varepsilon_{2\nu}^{(1)}
\int Dz \frac{q_{\lambda}}{Q^2 + i\varepsilon} \left[S_{21,\sigma}
+ \frac{q.S_{21}} {Q^2 + i\varepsilon} p_{z\sigma}
\right] F_5(z_+,z_-)~.
\end{eqnarray}

A similar structure is obtained for the {\sc Pauli}--type terms. The
unpolarized contributions read
\begin{eqnarray}
\label{TG11L}
\T^G_{11} &=&
 2  \overline{u}(p_2,S_2)
\sigma_{\alpha\beta} q^{\alpha} p_-^{\beta} u(p_1,S_1)
  \left[G_1(\xi,\eta) + \varepsilon_{1}^{(2)\mu} \varepsilon_{1}^{(1)\nu}
  G_{2,\mu\nu}(\xi,\eta)\right]\\
\label{TG22L}
\T^G_{22} &=&
2 \overline{u}(p_2,S_2)
\sigma_{\alpha\beta} q^{\alpha} p_-^{\beta} u(p_1,S_1)
  \left[G_1(\xi,\eta) + \varepsilon_{2}^{(2)\mu} \varepsilon_{2}^{(1)\nu}
  G_{2,\mu\nu}(\xi,\eta)\right] \\
\T^{G}_{kl} &\propto& \left(\frac{1}{\nu}\right)
^{1/2+n}~~~~~~{\rm for~the~other~projections}~~{k,l~\epsilon~\{1,2,3\}}~~
{\rm and}~~~{n \geq 0}~,
\end{eqnarray}
resulting into
\begin{eqnarray}
\T_{11}^G = \T_{22}^G~.
\end{eqnarray}
Here the tensor $G_{2,\mu\nu}$ is obtained from Eq.~(\ref{eqF2H})
substituting $F$ into $G$.

The polarized {\sc Pauli}--type terms obey~:
\begin{eqnarray}
\T_{12}^{G5} = -\T_{21}^{G5}
\end{eqnarray}
\begin{eqnarray}
\label{EQG5}
\T^{G5}_{kl} &\propto& \left(\frac{1}{\nu}\right)
^{1/2+n}~~~~~~{\rm for~the~other~projections}~~{k,l~\epsilon~\{1,2,3\}}~~
{\rm and}~~~{n \geq 0}~,
\end{eqnarray}
with
\begin{eqnarray}
\T^{G5}_{12} &=&  i~\varepsilon^{\mu\lambda\nu\sigma}
\varepsilon_{1}^{(2)\mu} \varepsilon_{2}^{(1)\nu} 
\int Dz \frac{q_{\lambda}}{Q^2 + i\varepsilon} \left[\Sigma_{21,\sigma}
+ \frac{q.\Sigma_{21}}{Q^2 + i\varepsilon} p_{z\sigma} \right]
G_5(z_+,z_-)~,
\end{eqnarray}
and
\begin{eqnarray}
\Sigma_{21}^{\sigma} := 
- \frac{1}{2} \overline{u}(p_2,S_2) \gamma_5 
\sigma^{\sigma\alpha} p_{-\alpha} u(p_1,S_1)~.
\end{eqnarray}
A comparison of
the structure of the projections $\T_{12(21)}^{F5,G5}$ with the
corresponding tensor structure for forward scattering, cf.~\cite{BGR,BT},
shows  that the two contributions are related to terms containing
the following contributions of structure functions
\begin{eqnarray}
\propto  q_{\lambda} S_{\sigma,21} &\rightarrow& g_1(x_B) +g_2(x_B) \\
\propto  q_{\lambda} p_{z\sigma}  &\rightarrow& g_2(x_B)~.
\end{eqnarray}

In summary, the non--forward  Compton amplitude at the level of the 
twist--2  contributions in lowest order QCD in the generalized Bjorken 
region is described by two helicity states of both virtual photons. 
The unpolarized and polarized {\sc Dirac}-- and {\sc Pauli}--terms
are related to two non--perturbative amplitude functions respectively.
\section{The Integral Relations}
\renewcommand{\theequation}{\thesection.\arabic{equation}}
\setcounter{equation}{0}
\label{sec-6}

\vspace{1mm}
\noindent
The amplitude functions describing the twist--2 contributions to the
virtual Compton amplitude in the generalized Bjorken region obey relations
to which we turn now.
\subsection{Unpolarized Contributions}
\renewcommand{\theequation}{\thesection.\arabic{equation}}
\setcounter{equation}{0}
\label{sec-6.1}

\vspace{1mm}
\noindent
For the unpolarized contributions to the Compton amplitude the relations
\begin{eqnarray}
\label{CGgen}
\T_{11}^{F,G} = \T_{22}^{F,G}
\end{eqnarray}
hold for the diagonal helicity projections. All other projections vanish
at the level of
the twist--2 contributions in lowest order QCD in the generalized
Bjorken region. 
Although
Eq.~(\ref{CGgen}) holds, the partonic interpretation of $\T_{11}^{F,G}$
and $\T_{22}^{F,G}$ for non--forward scattering as given in the case of
forward scattering by
\begin{eqnarray}
\label{CGfor}
F_2(x_B) = 2 x F_1(x_B) \equiv \sum_{q} e_q^2 x \left[q(x_B) +
\overline{q}(x_B)\right]
\end{eqnarray}
has still to be clarified since the representations (\ref{TF11L},
\ref{TF22L}, \ref{TG11L}, \ref{TG22L}) 
contain yet two types of $Dz$--integrals~(\ref{eqH1},\ref{eqF2H}),
\begin{eqnarray}
\label{EQH1}
{\sf H_1}(\xi,\eta) &=& \int Dz \frac{\nu}{Q^2+i\varepsilon} 
H(z_+,z_-) = - \int Dz \frac{H(z_+,z_-)}
{\xi + t - i \varepsilon} \\
\label{EQH2}
{\sf H_2}(\xi,\eta) &=& \int Dz \frac{\nu~q.p_z}{(Q^2+i\varepsilon)^2}
H(z_+,z_-) = \int Dz \frac{t H(z_+,z_-)}  
{(\xi + t - i \varepsilon)^2}~.
\end{eqnarray}
Changing the integration variables to
$(t,z_-)$ the integration over $z_-$ can be performed,
\begin{eqnarray}
\widehat{H}(t,\eta) &=&
\int_{z_-^{\rm min}}^{z_-^{\rm max}}dz_- H(t-\eta z_-, z_-)
                    = \int_0^1 \frac{d\lambda}{\lambda^2} 
\int_{z_-^{\rm min}}^{z_-^{\rm max}} d z_-
                     h\left(\frac{t}{\lambda} -
                    \eta \frac{z_-}{\lambda}, \frac{z_-}{\lambda}
                    \right)   \nonumber\\
           &=& \int_t^{{\rm sign}(t)} \frac{dz}{z} \hat{h}(z,\eta)~,
\end{eqnarray}
with
\begin{eqnarray}
z_-^{\rm min, max} = \frac{t \pm 1}{\eta \pm 1}~,
\end{eqnarray}
and
\begin{eqnarray}
\label{hath}
\hat{h}(z,\eta) =
\int_{\rho_{\rm min}}^{\rho_{\rm max}} 
d\rho~h\left(z - \eta \rho,\rho\right)~,
\end{eqnarray}
where $z=t/\lambda, \rho=z_-/\lambda$ and
$\rho_{\rm min(max)} = z_{-,\rm min(max)}\cdot(z/t)$.
As it
will turn out below  $\hat{h}(z,\eta)$ denotes an amplitude
function on the partonic level. This interpretation does not apply to
$\widehat{H}(t,\eta)$.

By partial integration one may rewrite Eq.~(\ref{EQH2}).
The support of the variable $t$ is $t~\epsilon~[-1,+1]$. One obtains
\begin{eqnarray}
\label{eqpari}
\int_{-1}^{+1} dt \frac{t}{(\xi+t-i\varepsilon)^2} 
\int_t^{{\rm sign}(t)} \frac{dz}
{z} \hat{h}(z,\eta)
&=&
\int_{-1}^{+1} dt \frac{1}{\xi+t-i\varepsilon} 
\int_t^{{\rm sign}(t)} \frac{dz}{z} \hat{h}(z,\eta) \nonumber\\ & &
-
\int_{-1}^{+1} dt \frac{1}{\xi+t-i\varepsilon} \hat{h}(t,\eta)~,
\end{eqnarray}
which yields
\begin{eqnarray}
{\sf H_2}(\xi,\eta) = - {\sf H_1}(\xi,\eta) - \int_{-1}^{+1} dt
\frac{\hat{h}(t,\eta)}{\xi+t - i\varepsilon}~.
\end{eqnarray}
The $z$--integral contributions in 
Eqs.~(\ref{TF11L},\ref{TF22L},\ref{TG11L},\ref{TG22L})
cancel and  the helicity projections $T_{11(22)}^{F,G}$ are
\begin{eqnarray}
\label{par1}
\T_{11(22)}^{H}(\xi,\eta)
\propto - \int_{-1}^{+1}
\frac{\hat{h}(t,\eta)}{\xi+t - i\varepsilon} = - {\sf P} \int_{-1}^{+1}
dt \frac{\hat{h}(t,\eta)}{\xi+t} - i\pi \hat{h}(\xi,\eta)~,
\end{eqnarray}
which shows that $\T_{11(22)}^{H}$ obeys a `partonic' description. 
For the
{\sc Dirac}--type terms the
function $\hat{h}(\xi,\eta) = \hat{f}(\xi,\eta)$    turns 
into the quark and antiquark densities in the
forward limit $\eta \rightarrow 0, \xi \rightarrow x_B$.

To derive the relations on the Lorentz level, we consider 
Eq.~(\ref{comp5}). The unpolarized part of $T_{\mu\nu}$ contains the
following functions~:
\begin{eqnarray}
\label{hi1}
H_1(\xi,\eta) &=& - \frac{1}{\nu}
\int_{-1}^{+1} dt \frac{1}{\xi+t-i\varepsilon}
{\sf \hat{H}}(t,\eta) \\
\label{hi2}
H_k^{\sigma}(\xi,\eta) &=&
\frac{p_+^\sigma}{(-\nu)^k}
\int_{-1}^{+1} dt \frac{t}
{(\xi+t-i\varepsilon)^k}
{\sf \hat{H}}(t,\eta) +
\frac{\pi^\sigma}{(-\nu)^k}
\int_{-1}^{+1} dt \frac{1}
{(\xi+t-i\varepsilon)^2}
{\sf \tilde{H}_1}(t,\eta)
\\
\label{hi3}
H_{2,\mu\nu}(\xi,\eta) &=& \frac{1}{\nu^2} \int_{-1}^{+1} dt \frac{1}
{(\xi+t-i\varepsilon)^2} \Biggl\{
\left[-p_{+\mu}p_{+\nu} t^2 +(q_\nu p_{+\mu}
+q_\mu p_{+\nu}) t - g_{\mu\nu} q.p_+ t\right]
{\sf \hat{H}}(t,\eta) \nonumber\\ & &
+ \left[(p_{+\nu} \pi_\mu + p_{+\mu} \pi_{\nu}) t + (q_\nu \pi_\mu +
q_\mu \pi_\nu) \right] {\sf \tilde{H}_1}(t,\eta)
- \pi_{\mu} \pi_{\nu} {\sf \tilde{H}_2}(t,\eta)\Biggr\}~,
\end{eqnarray}
where $k = 1,2$ and
\begin{eqnarray}
\label{eqhat}
{\sf \hat{H}}(t,\eta) &=& \int_t^{{\rm sign}(t)} \frac{dz}{z} 
\hat{h}(z,\eta) \\
\label{eqtilde}
{\sf \tilde{H}_k}(t,\eta) &=& \int_t^{{\rm sign}(t)} \frac{dz}{z}
\tilde{h}_k(z,t,\eta)~,
\end{eqnarray}
with
\begin{eqnarray}
\tilde{h}_k(z,t,\eta) = \left(\frac{t}{z}\right)^k
\int_{\rho_{\rm min}}^{\rho_{\rm max}} d\rho~\rho^k~h(z-\eta\rho,\rho)~.
\end{eqnarray}
Note that contractions of the vector $\pi_{\sigma}$ in the above 
equations
with one of the vectors $q^\mu, p_\pm^\mu$ or $n_2^\mu$, which span the
 Minkowski space, yields contributions of at most
$O(\mu^2)$ if compared to the large invariants which are of $O(\nu)$.
The contributions due to these terms are therefore hadronic mass scale
corrections or of higher twist. These terms vanish also both for forward
scattering and in the case of vacuum--meson transition 
$p_+ = p_-, \eta=1$. Due to this we will consider only the terms which
do not contain the vectors $\pi_{\sigma}$ in the following. The integrals
(\ref{hi2},\ref{hi3}) may be rewritten by partial integration.
\begin{eqnarray}
\label{hi2a}
H_2^{\sigma}(\xi,\eta) &=&
\frac{p_+^\sigma}{\nu^2}
\int_{-1}^{+1} dt \frac{1}
{\xi+t-i\varepsilon}\left[
{\sf \hat{H}}(t,\eta) - \hat{h}(t,\eta)\right] + O(\pi^{\sigma})
\\
\label{hi3a}
H_{2,\mu\nu}(\xi,\eta) &=& \frac{1}{\nu^2} \int_{-1}^{+1} dt \frac{1}
{\xi+t-i\varepsilon} \Biggl\{-p_{+\mu} p_{+\nu} t \left[2
{\sf \hat{H}}(t,\eta) - \hat{h}(t,\eta) \right] \nonumber\\
 & &~~~~~~~~~
+
\left[(q_\nu p_{+\mu}
+q_\mu p_{+\nu})   - g_{\mu\nu} q.p_+  \right]\left[
{\sf \hat{H}}(t,\eta) - \hat{h}(t,\eta) \right] \Biggr\} 
+ O(\pi_{\mu(\nu)})
\end{eqnarray}
Let us define the vector
\begin{eqnarray}
P_{21}^{\sigma} := \overline{u}(p_2,S_2) \gamma^{\sigma} u(p_1,S_1),
\end{eqnarray}
with $P_{21}^{\sigma} = p_+^{\sigma}$ for forward scattering. For the
unpolarized part of $T^{\mu\nu}$,~Eq.~(\ref{comp5}), $A^{\mu\nu}$,
one obtains
\begin{eqnarray}
\label{eqAmn}
A^{\mu\nu} &=&  -2
\frac{q.P_{21}}{\nu} 
\left[g^{\mu\nu} - \frac{q^\mu p_+^\nu + q^\nu p_+^\mu }{q.p_+}\right] 
\int_{-1}^{+1} dt \frac{{\sf F_1}(t,\eta)}{\xi + t -i\varepsilon} 
\nonumber\\ & &
+ \frac{2}{\nu} \left[q^\mu\left(P_{21}^\nu - p_+^\nu \frac{q.P_{21}}
{\nu}\right) +   q^\nu \left(P_{21}^\mu - p_+^\mu \frac{q.P_{21}}
{\nu}\right) \right]
\int_{-1}^{+1} dt
\frac{{\sf \hat{H}}(t,\eta)}{\xi + t -i\varepsilon}~,
\nonumber\\ & &
- \frac{q.P_{21}}{\nu^2}       p_+^\mu p_+^\nu     \int_{-1}^{+1}
dt \frac{{\sf F_2}(t,\eta)}{\xi + t - i \varepsilon}
\nonumber\\ & &
- \frac{2}{\nu} \left[p^\mu_+\left(p_{21}^\nu - p_+^\nu \frac{q.P_{21}}
{\nu}\right) +
                      p^\nu_+\left(p_{21}^\mu - p_+^\mu \frac{q.P_{21}}
{\nu}\right) \right]
\int_{-1}^{+1} dt
\frac{t~{\sf \hat{H}}(t,\eta)}{\xi + t -i\varepsilon}~.
\end{eqnarray}
Here we defined
\begin{eqnarray}
\label{eqF1}
{\sf F_1}(t,\eta) &=& \hat{h}(t,\eta) \\
\label{eqF2}
{\sf F_2}(t,\eta) &=& 2 t~\hat{h}(t,\eta)~.
\end{eqnarray}
The vector
\begin{eqnarray}
\Pi^{\mu} = P_{21}^{\mu} - p_+^\mu \frac{q.P_{21}}{\nu}
\end{eqnarray}
in Eq.~(\ref{eqAmn}), as also the case for the vector $\pi^{\mu}$ above,
has contractions with the vectors $q^\mu, p_\pm^\mu$ and $n_2^\mu$
which either vanish or are of $O(\mu^2)$ only. Therefore these 
terms         do not contribute at the lowest twist level unlike those
due to ${\sf F_1}(t,\eta)$ and ${\sf F_2}(t,\eta)$.
\begin{eqnarray}
{\sf F_2}(t,\eta) = 2 t~{\sf F_1}(t,\eta)
\end{eqnarray}
is the generalization of the {\sc Callan}--{\sc Gross} relation for
non--forward scattering. These distribution amplitudes have the partonic
representation~(\ref{eqF1}, \ref{eqF2}), whereas the other distribution
amplitudes of non-leading twist in Eq.~(\ref{eqAmn}) 
depend on the function
${\sf \hat{H}}
(t,\eta)$, which is related to $\hat{h}(t,\eta)$ by the
integral~Eq.~(\ref{eqhat}).
\subsection{Polarized Contributions}
\renewcommand{\theequation}{\thesection.\arabic{equation}}
\label{sec-6.2}

\vspace{1mm}
\noindent
The matrix element ${\sf T}_{12}^{H5}$, Eq.~(\ref{eqpol}),
\begin{eqnarray}
{\sf T}_{12}^{H5} = i~\varepsilon^{\mu\lambda\nu\sigma}
\varepsilon_{1\mu}^{(2)}
\varepsilon_{2\nu}^{(1)} B_{\lambda\sigma}
\end{eqnarray}
contains the tensor
\begin{eqnarray}
B_{\lambda\sigma} =
\int Dz \frac{q_{\lambda}}{Q^2 + i\varepsilon} \left[S^H_{21,\sigma}
+ \frac{q.S^H_{21}}{Q^2 + i\varepsilon} p_{z\sigma} \right]
H_5(z_+,z_-), \nonumber
\end{eqnarray}
with  $S^H_{21} = S_{21} (\Sigma_{21})$
for $H = F(G)$. It may be rewritten as
\begin{eqnarray}
\label{eqbls1}
B_{\lambda\sigma} =
- \frac{1}{\nu}
\int Dz \frac{q_{\lambda}}{\xi + t - i\varepsilon} \left[S^H_{21,\sigma}
- \frac{1}{\nu}
\frac{t~q.S^H_{21}}{\xi + t - i\varepsilon} p_{+\sigma}
 + \frac{1}{\nu} \frac{q.S^H_{21}}{\xi + t - i \varepsilon}
z_- \pi_{\sigma} \right] H_5(z_+,z_-)~. \nonumber\\
\end{eqnarray}
The latter term in Eq.~(\ref{eqbls1})
vanishes for forward scattering and in the case
of vacuum--meson transition. We perform the integration over
$z_-$ and obtain
\begin{eqnarray}
\label{Beq}
B_{\lambda\sigma} &=& - \frac{1}{\nu} q_{\lambda} S^H_{21,\sigma}
\int_{-1}^{+1} dt \frac{1}{\xi+t-i\varepsilon} \int_t^{{\rm sign}(t)}
\frac{dz}{z} \hat{h}_5(z,\eta) \nonumber\\  & &
- \frac{1}{\nu^2} q_{\lambda} p_{+\sigma}~q.S^H_{21} \int_{-1}^{+1}
dt \frac{1}{\xi + t - i \varepsilon} \left[\hat{h}_5(t,\eta) -
\int_t^{{\rm sign}(t)} \frac{dz}{z} \hat{h}_5(z,\eta) \right]
\nonumber\\ & &
- \frac{1}{\nu^2} q_{\lambda} \pi_{\sigma}~q.S^H_{21}
\int_{-1}^{+1} dt
\frac{1}{\xi + t - i\varepsilon} \int_t^{{\rm sign}(t)} \frac{dz}{z}
\widetilde{h}_5(z,t,\eta)~.
\end{eqnarray}
Here the distribution amplitude $\widetilde{h}_5$ denotes the first
moment with respect to $z_-$ of the function $H_5(z_+,z_-)$
\begin{eqnarray}
\label{htil}
\widetilde{h}_5(z,t,\eta) = \left(\frac{t}{z}\right)
\int_{\rho_{\rm min}}^{\rho_{\rm max}} 
d\rho~\rho h\left(z - \eta \rho,\rho\right)~,
\end{eqnarray}
whereas $\hat{h}(z,\eta)$, Eq.~(\ref{hath}), is the
corresponding 0th moment.

Let us rewrite Eq.~(\ref{Beq}) choosing the notation in analogy to the
forward case by
\begin{eqnarray}
\label{Beq1}
\lefteqn{
B_{\lambda\sigma} = - \frac{1}{\nu} q_{\lambda} \int_{-1}^{+1} 
\frac{dt}{\xi + t - i\varepsilon}} \nonumber\\  & &
\times
\left\{
S^H_{21,\sigma} \left[
{\sf G_1}(t,\eta) +
{\sf G_2}(t,\eta) \right]
+ \frac{1}{\nu} p_{+\sigma}~q.S^H_{21}
{\sf G_2}(t,\eta)
+ \frac{1}{\nu}  \pi_{\sigma}~q.S^H_{21}
{\sf G_3}(t,\eta) \right\}~.
\end{eqnarray}
One obtains
\begin{eqnarray}
\label{par2}
{\sf G_1}(t,\eta) &:=& \hat{h}_5(t,\eta)\\
\label{WWNEW}
{\sf G_2}(t,\eta) &=& - {\sf G_1}(t,\eta) + \int_{t}^{{\rm sign}(t)}
\frac{dz}{z}{\sf G_1}(z,\eta)\\
\label{SFnew}
{\sf G_3}(t,\eta) &:=&  \int_t^{{\rm sign}(t)} \frac{dz}{z}
\widetilde{h}_5(z,t,\eta)~.
\end{eqnarray}
The $t$--integrals in Eq.~(\ref{Beq1}) are performed further according to
\begin{eqnarray}
\int_{-1}^{+1} dt \frac{A(t)}{\xi+t-i\varepsilon} =
{\sf P} \int_{-1}^{+1} dt \frac{A(t)}{\xi+t} + i \pi A(-\xi)~.
\end{eqnarray}

The polarized non--forward
distribution amplitude ${\sf G_1}(t,\eta)$ has a partonic interpretation
as in the unpolarized
case~Eq.~(\ref{par1}). Eq.~(\ref{WWNEW})
is the non--forward generalization of the {\sc Wandzura}--{\sc Wilczek}
relation. For the special case of vacuum--meson transition according
integral relations were discussed in~\cite{BBKT}.
In Eq.~(\ref{SFnew})  a {\it new} distribution amplitude 
${\sf G_3}(t,\eta)$ emerges  which is, however, not of twist--2  since
the contractions of $\pi_{\sigma}$ with the       4--momenta of the
scattering process are of $O(\mu^2)$ if compared to the other terms
which are of $O(\nu)$.

The above results show that the well--known results for forward
scattering, the {\sc Callan}-{\sc Gross} and the {\sc Wandzura}--{\sc
Wilczek} relation are not bound to forward scattering only, but are
of more general validity. In particular their derivation does not 
require
to use the optical theorem. Furthermore, the above relations hold for
quite general functions $H_{(5)}(z_+,z_-)$. In deriving the above
relations assumptions on their complex structure had not to be made.
The $O(\mu^2/\nu)$ corrections are associated with new distribution 
amplitudes,
cf. the contributions $\propto \pi_{\sigma}$ or $\Pi_{\sigma}$ 
in Eqs.~(\ref{eqH1},
\ref{eqH2}, \ref{hi2}, \ref{hi3}, \ref{eqAmn},
\ref{Beq1}), which are either representable by integral
relations of leading twist distribution amplitudes or are
higher moments in $z_-$
of the functions $H(z_+,z_-)$ and $H^5(z_+,z_-)$, respectively.
\subsection{Forward Scattering}
\renewcommand{\theequation}{\thesection.\arabic{equation}}
\label{sec-6.3}
After the above considerations the limit to forward scattering is easily
performed. Because the {\sc Pauli}--type terms vanish linearly with
$p_-$ only the {\sc Dirac}--type terms remain.  For forward scattering 
both the functions ${\sf F_1}(\xi,\eta)$ and ${\sf G_1}(\xi,\eta)$ are
real. The absorptive part of the Compton amplitude, the hadronic tensor
$W_{\mu\nu}$, is given by
\begin{eqnarray}
W_{\mu\nu} = \frac{1}{2\pi} {\sf Im}~T_{\mu\nu}~.
\end{eqnarray}
To derive the forward structure functions
we rewrite
Eqs.~(\ref{eqAmn}, \ref{Beq}) applying the symmetry relations
Eqs.~(\ref{symrel1}, \ref{symrel2}) interchanging $(z_+,z_-)
\leftrightarrow (-z_+,-z_-)$, i.e. $t \leftrightarrow -t$, respectively.
We define the branches of the functions ${\sf F_1}(t,0)$
and ${\sf G_1}(t,0)$ for $-1 \leq t < 0$ and $0 < t \leq +1$ as
the antiquark and quark distribution function     by
\begin{eqnarray}
\label{qdis1}
{\sf F_1}(t,0) &=&              \sum_q~e_q^2 \left[
q(t) \theta(t) 
                  -  \overline{q}(-t) \theta(-t)\right] \\
\label{qdis2}
{\sf G_1}(t,0) &=&              \sum_q~e_q^2 \left[
\Delta q(t) \theta(t) 
                  +  \Delta \overline{q}(-t) \theta(-t)\right]~.
\end{eqnarray}
For the unpolarized contributions one obtains for forward scattering,
 $q = q_1 = q_2,~p=p_+/2$,
\begin{eqnarray}
q_{\mu} A^{\mu\nu}     =             p^\nu \left[
\int_{-1}^{+1}
\frac{2\xi {\sf F_1}(t,0) - {\sf F_2}(t,0)}{\xi-t - i\varepsilon}
- \int_{-1}^{+1}
\frac{2\xi {\sf F_1}(t,0) + {\sf F_2}(t,0)}{\xi+t - i\varepsilon}
\right]  = 0~.
\end{eqnarray}
Taking the absorptive part yields
\begin{eqnarray}
\pm 2 \xi {\sf F_1}(\pm \xi,0) = {\sf F_2}(\pm \xi,0)~.
\end{eqnarray}
The structure functions are now given by
\begin{eqnarray}
F_1(x_B) &=& \frac{1}{2} \left[{\sf F_1}(\xi,0) - {\sf F_1}(-\xi,0)\right]
= \frac{1}{2} \sum_q e_q^2 \left[q(x_B) + \overline{q}(x_B)\right]~,
\\
F_2(x_B) &=& {\sf F_2}(\xi,0) + {\sf F_2}(-\xi,0)~,
\end{eqnarray}
with the Bjorken variable $x_B = \lim_{\eta \rightarrow 0} \xi$. The
structure functions obey
\begin{eqnarray}
F_2(x_B) &=& 2 x_B F_1(x_B)
\end{eqnarray}
the {\sc Callan}--{\sc Gross} relation~\cite{CG}.

Correspondingly, for the polarized part one obtains
\begin{eqnarray}
\label{Beqa}
B_{\lambda\sigma} &=& - \frac{1}{2\nu} q_{\lambda} S^H_{21,\sigma}
\int_{-1}^{+1} dt \left[
\frac{{\sf G_1}(t,0)+{\sf G_2}(t,0)}{\xi+t-i\varepsilon} +
\frac{{\sf G_1}(t,0)+{\sf G_2}(t,0)}{\xi-t-i\varepsilon} \right]
\\  & &
- \frac{1}{2\nu^2} q_{\lambda} p_{+\sigma}~q.S^H_{21} \int_{-1}^{+1}
dt \left[
\frac{{\sf G_2}(t,0)}{\xi + t - i \varepsilon} +
\frac{{\sf G_2}(t,0)}{\xi - t - i \varepsilon} \right]~.
\nonumber
\end{eqnarray}
The absorptive part is described by the  structure functions
$g_1(x_B)$ and $g_2(x_B)$ with, cf.~(\ref{Beq}),
\begin{eqnarray}
\label{eqg1}
g_1(x_B) &=& \frac{1}{2} \left[{\sf G_1}(\xi,0) + {\sf G_1}(-\xi,0)\right]
= \frac{1}{2} \sum_q e_q^2 \left[\Delta q(x_B) 
+ \Delta \overline{q}(x_B)\right]  \\
\label{eqg2}
g_2(x_B)       &=& - g_1(x_B)       + \int_{x_B}^{1}
\frac{dz}{z} g_1(z)~.
\end{eqnarray}
Eq.~(\ref{eqg2}) is the {\sc Wandzura}--{\sc Wilczek} relation\footnote{
In the presence of electroweak currents five polarized structure functions
contribute on the level of twist-2 and for the quark operators
also for twist--3. These structure functions are connected by two
further relations for the twist--2 contributions, the
{\sc Dicus}-relation~\cite{DIC} and a relation by {\sc Bl\"umlein} and 
{\sc Kochelev}~\cite{BK}. The three twist--3 relations among the
respective contributions to the polarized structure functions due to the
quark operators at lowest order QCD have been derived in Ref.~\cite{BT} 
recently. One of these relations applies to the case of pure 
electromagnetic interactions.}~\cite{WW}.
\section{Conclusions}
\renewcommand{\theequation}{\thesection.\arabic{equation}}
\setcounter{equation}{0}
\label{sec-8}
We studied the structure of the virtual Compton amplitude for
deep--inelastic non--forward scattering $\gamma^* + p \rightarrow
\gamma^{(*)'} +p'$ in lowest order in QCD in the massless limit. In the
generalized Bjorken region $q.p_+, -q^2 \rightarrow \infty$ the twist--2 
contributions to the Compton amplitude were calculated using the 
non--local operator product expansion for general spin states. In this 
approximation the Compton amplitude consists of an unpolarized and a 
polarized {\sc Dirac}-- and {\sc Pauli}--type amplitude, the latter of 
which vanishes in the case of forward scattering. The expectation values 
of the (non--local) twist--2 vector operators in the non--forward case do
still contain terms $\propto 1/\nu^{1/2+k},~k \geq 0$, which are 
contributions of twist--3 and higher to the Compton amplitude. 
These contributions have to be considered in common with the non--forward
expectation values of the higher twist operators.
A decomposition of the 
amplitude was performed with respect to the helicity states of both 
(virtual) photons. The twist--2 contributions are due to two polarization
states only, which are for the unpolarized part $\T_{11}$ and $\T_{22}$ 
and for the polarized part $\T_{12}$ and $\T_{21}$. For the twist--2
contributions the gauge invariance of the non--local light cone expansion
was proven in the non--forward case in  the generalized  Bjorken region.
The relations between the twist--2 contributions of the unpolarized and 
polarized amplitude functions were derived. They are the non--forward 
generalizations of the {\sc Callan}--{\sc Gross} and 
{\sc Wandzura}--{\sc Wilczek} relations for unpolarized and polarized 
deep--inelastic forward scattering. The relations for the {\sc Dirac} and
{\sc Pauli} parts are of the same form.

\vspace{1mm}
\noindent
{\bf Acknowledgement}\\
Our thanks are due to B. Geyer for fruitful discussions and U.~Gensch for
his constant support. Discussions with A. Tkabladze  in an early phase of 
this project are acknowledged. J.B. would like to thank the Institute of 
Theoretical Physics at Graz University for their kind hospitality. D.R. 
likes to thank DESY Zeuthen and the Institute of Theoretical Physics at 
Graz University for the kind hospitality extended to him, in particular
to C.B. Lang, H. Mitter, N. Pucker, and W.~Schweiger.
\newpage

\end{document}